\documentclass[aps,prb,twocolumn,showpacs,superscriptaddress]{revtex4-1}

\usepackage{graphicx}
\usepackage{dcolumn}
\usepackage{bm}
\usepackage{amssymb} 
\usepackage{mhchem}
\usepackage{amsmath}
\usepackage{xcolor} 
\usepackage{makecell} 
\begin{document}

\title[Thermal boundary resistance from transient nanocalorimetry: a multiscale modeling approach]{Thermal boundary resistance from transient nanocalorimetry: a multiscale modeling approach} 

\author{Claudia Caddeo}
\affiliation{Dipartimento di Matematica e Fisica, Universit\`a Cattolica del Sacro Cuore, Via Musei 41, I-25121 Brescia, Italy}
\affiliation{Interdisciplinary Laboratories for Advanced Materials Physics (I-LAMP), Universit\`a Cattolica del Sacro Cuore, Via Musei 41, I-25121 Brescia, Italy}
%
 
\author{Claudio Melis}%
 \affiliation{Dipartimento di Fisica, Universit\`a degli Studi di Cagliari, Cittadella Universitaria, I-09042 Monserrato (Ca), Italy}

\author{Andrea Ronchi}
\affiliation{Dipartimento di Matematica e Fisica, Universit\`a Cattolica del Sacro Cuore, Via Musei 41, I-25121 Brescia, Italy}
\affiliation{Interdisciplinary Laboratories for Advanced Materials Physics (I-LAMP), Universit\`a Cattolica del Sacro Cuore, Via Musei 41, I-25121 Brescia, Italy}
\affiliation{Laboratory of Solid State Physics and Magnetism, Department of Physics and Astronomy, KU Leuven, Celestijnenlaan 200D, B-3001 Leuven, Belgium}

\author{Claudio Giannetti}
\affiliation{Dipartimento di Matematica e Fisica, Universit\`a Cattolica del Sacro Cuore, Via Musei 41, I-25121 Brescia, Italy}
\affiliation{Interdisciplinary Laboratories for Advanced Materials Physics (I-LAMP), Universit\`a Cattolica del Sacro Cuore, Via Musei 41, I-25121 Brescia, Italy}

\author{Gabriele Ferrini}
\affiliation{Dipartimento di Matematica e Fisica, Universit\`a Cattolica del Sacro Cuore, Via Musei 41, I-25121 Brescia, Italy}
\affiliation{Interdisciplinary Laboratories for Advanced Materials Physics (I-LAMP), Universit\`a Cattolica del Sacro Cuore, Via Musei 41, I-25121 Brescia, Italy}

\author{Riccardo Rurali}%
\affiliation{Institut de Ci\`encia de Materials de Barcelona (ICMAB-CSIC), Campus de Bellaterra, 08193 Bellaterra, Barcelona, Spain
}%

\author{Luciano Colombo}
\affiliation{Dipartimento di Fisica, Universit\`a degli Studi di Cagliari, Cittadella Universitaria, I-09042 Monserrato (Ca), Italy}%
%

\author{Francesco Banfi}
\affiliation{Dipartimento di Matematica e Fisica, Universit\`a Cattolica del Sacro Cuore, Via Musei 41, I-25121 Brescia, Italy}
\affiliation{Interdisciplinary Laboratories for Advanced Materials Physics (I-LAMP), Universit\`a Cattolica del Sacro Cuore, Via Musei 41, I-25121 Brescia, Italy}
%
             
\begin{abstract}
The Thermal Boundary Resistance at the interface between a nanosized Al film and an Al$_2$O$_3$ substrate is investigated at an atomistic level.
A room temperature value of 1.4 m$^{2}K/GW$ is found. The thermal dynamics occurring in time-resolved thermo-reflectance experiments is then modelled via macro-physics equations upon insertion of the materials parameters obtained from atomistic simulations. Electrons and phonons non-equilibrium and spatio-temporal temperatures inhomogeneities are found to persist up to the nanosecond time scale. These results question the validity of the commonly adopted lumped thermal capacitance model in interpreting transient nanocalorimetry experiments. The strategy adopted in the literature to extract the Thermal Boundary Resistance from transient reflectivity traces is revised at the light of the present findings. The results are of relevance beyond the specific system, the physical picture being general and readily extendable to other heterojunctions.
\end{abstract}


\maketitle

%

\section{Introduction}
Heat transfer at the meso/nano-scale represents an outstanding challenge, among the most relevant under an applicative standpoint \cite{Chen2005,luo2013nanoscale,cahill2014nanoscale,hoogeboom2015}.  
In this context Thermal Boundary Resistance (TBR), the parameter ruling heat transfer at the interface between two materials\cite{swartz1999}, plays a key role. Accessing the TBR between nanosized metals and insulating substrates remains an open issue and the prerequisite to enhance heat dissipation in next-generation micro and nano-devices. Indeed low thermal dissipation across heterojunctions is among the main impediments towards further circuits downscaling \cite{moore2014,pop2010}.

Much effort has been devoted to access the TBR at metal-insulator interfaces both theoretically and experimentally\cite{cahill2014nanoscale,Chen2005,landry2009}. Unfortunately theoretical predictions deviate from TBR values extracted from time-resolved thermo reflectance (TR-TR) measurements\cite{costescu2003,majumdar2004}, the go-to technique to inspect TBR at these junctions\cite{stoner1993,cahill2003nanoscale,Siemens2010,juve2009cooling,Banfi_APL2012,stoll2015}. The present work, based on multi-scale modelling, reconciles the discrepancy for the paradigmatic case of the Al-Sapphire heterojunction.

The basic idea to extract the TBR from a metal-dielectric interface via a TR-TR experiment is as follows. A short laser pump pulse delivers energy to the metallic film, triggering an impulsive temperature dynamics. The temperature dynamics (a) is ruled by the materials thermal parameters, among which the TBR, and (b) affects the temperature-dependent optical constants of the sample resulting in a transient reflectivity variation. The latter is investigated by means of time-delayed laser¡Çs probe pulses, the time-delay being with respect to the excitation instant. The information on the TBR is ultimately encoded in the sample's transient reflectivity changes. 
In order to retrieve the TBR a model linking the TBR to the thermal dynamics is thus required. 

Typically the metal film is modelled as a lumped thermal capacitance\cite{ozisik1993heat} exchanging heat with the underlying dielectric substrate through a TBR. This implies a unique time-varying film temperature (electrons and phonons anchored at the same temperature), constant throughout the film thickness. The reflectivity variation is then linked to the film's spatially homogenous temperature profile.

In the present work we challenge the validity of this model for values of the TBR that might be expected at solid-solid interfaces and argue that its application in the fitting procedure of the time-resolved reflectivity traces leads to an overestimation of the true TBR. We adopt multi-scale modeling to rationalize the thermal dynamics occurring in TR-TR measurements. On this basis a procedure to extract the TBR from transients experiments is proposed.

The present approach is based on atomistic modeling of the TBR and the subsequent description of the impulsive thermodynamics beyond the lumped thermal capacitance approach. The onset of spatio-temporal temperatures gradients in the metal film during all-optical time-resolved nanocalorimetry experiments is shown. TR-TR experiments probe a temperature dynamics - taking place in proximity of the metal film surface - other than the one controlling the TBR - occurring at the film interface. Based on this evidence a strategy is put forward to extract the TBR from TR-TR traces. The emerging physical picture and the TBR retrieval protocol are rather general and may find application in a variety of interfaced systems in addition to the present one. 

The work is organized following a bottom-up progression both in dimensions and time scales. In Section \ref{Why the Al-Al$_{2}$O$_{3}$ heterojunction} the suitability of the Al-Al$_{2}$O$_{3}$ heterojunction is briefly addressed. In Section \ref{MD} the material¡Çs thermal parameters - TBR and lattice thermal conductivities - are calculated at an atomistic level by Non-Equilibrium Molecular Dynamics (NEMD). Theoretical assesment of the TBR is of paramount importance since its value influences the impulsive thermal dynamics occurring in TR-TR experiments. Attention is therefore devoted to validate the results by inspecting the temperature dependence of the TBR and verifying it¡Çs truly interface character. We pinpoint that if, instead of the theoretically calculated TBR values, we were to rely on the experimental values of the TBR as retrieved from previous TR-TR measurements (that is obtained by fitting the experimental data with a lamped thermal capacitance model) we would run the risk of forcing the temperature distribution to be spatially homogeneous, conformally to the ansatz behind the fitting model. This concept is discussed in Section \ref{Detection process: TBR from Time Resolved Thermo-Reflectance}.
Section \ref{Transient nanocalorimetry} addresses the thermal dynamics occurring in TR-TR experiment via continuous macrophysics equations (subsection \ref{Excitation process: impulsive thermodynamics}) and its link to both the transient thermo-reflectivity traces and the TBR retival process (subsection \ref{Detection process: TBR from Time Resolved Thermo-Reflectance}). The modelling, based on Finite Element Methods (FEM), involves the interplay of Maxwell equations, the two-temperature model (TTM) and Fourier law for heat transport with insertion of the material¡Çs thermal parameters calculated from NEMD. The implication of the thermal dynamics on TR-TR experiments, and on the TBR therein accessed, are theoretically discussed. An effective TBR retrieval protocol is proposed, readily extendable to other interfaces. Finally (subsection \ref{Steady-state VS transient nanocalorimetry}), the present findings are discussed at the light of recent literature¡Çs results in the frame of $\textit{Steady-state}$ heat transfer.

\section{Why the A\MakeLowercase{l}-A\MakeLowercase{l}$_{2}$O$_{3}$ heterojunction}
\label{Why the Al-Al$_{2}$O$_{3}$ heterojunction} 
The Al-Al$_{2}$O$_{3}$ interface represents an ideal model system to address the issue at hand. Sapphire is optically transparent whereas Al absorbs energy over a wide spectral range - UV to the IR - allowing for selective heating of the metal film only upon laser pulse absorption. Sapphire has no porosity nor grain boundaries, thus allowing for a well defined interface, a key feature needed to minimize the possible influence of roughness-driven phonon scattering pathways on the TBR. For this materials combination, the effect of direct electron-phonon coupling across the interface on the TBR may be disregarded \cite{Lombard2014}. Experimental TBR data, as retrieved from TR-TR measurements, are available in the literature\cite{stoner1993}. These aspects allow focusing on the relevant physics avoiding unnecessary complications. Nevertheless, the emerging physical picture is rather general and may apply to heterojunctions other than the present one.

Assessment of the thermal parameters in this system has an immediate technological fall out. The TBR between an Al thin film and the supporting sapphire is of relevance in a variety of applications ranging from broad-band optical mirrors\cite{Chkhalo1995} to high frequency optoacoustic transducers\cite{nardiAPL2012} and protective coatings technology\cite{ksiazek2002}. 

\section{Materials thermal parameters from atomistic modelling}
In this section the materials thermal parameters - TBR and lattice thermal conductivities - are calculated at an atomistic level on a system of realistic size. The outcome will then serve as the input for modelling the impulsive thermodynamics taking place in TR-TR experiments as addressed in Section \ref{Transient nanocalorimetry}. 
\label{MD}
\subsection{Non-Equilibrium Molecular Dynamics}
\label{NEMD}
\subsubsection{Computational setup}
\label{Computational setup} 
All simulations have been performed with the LAMMPS package \cite{LAMMPS}. Among others\cite{lazic2012,scopece2015} we selected the Streitz-Mintmire potential for Al and Al$_2$O$_3$, where the total energy is given by the sum of the electrostatic energy and a standard embedded atom method potential (EAM)\cite{streitz1994,zhou2004}.
The potential allows a dynamical self-consistent calculation of the atomic partial charges to take into account the charge transfer between the oxide and the metal at the interface. Electrostatic interactions have been calculated with the Wolf summation method\cite{wolf1999}, with cutoff $\rho$=8\AA\  and damping parameter $\alpha$=0.11\AA$^{-1}$.
The adopted potential has been tested against several properties of Al and Al$_{2}$O$_{3}$ (e.g. lattice constants, cohesive energies and elastic constants), providing a very good agreement with experimental and ab initio data\cite{streitz1994,zhou2004}.

NEMD\footnote{Refer to Ref.~[\!\!\citenum{Dettori2016}] for a review of NEMD and a thorough discussion on different formalisms avalable for calculating TBR.} has been used to calculate the lattice thermal conductivity of the pure Al thin film, of the pure Al$_2$O$_3$ substrate and to determine the TBR, $R_{ph}$. The subscript recalls that the TBR in this system is solely determined by phonons, as will be addressed further on. The temperature was controlled by a Nos\'e-Hoover thermostat while the equations of motion have been integrated by a time-step as short as 1 fs. 
In brief, NEMD consists in coupling the opposite ends of the sample to two thermostats at different temperature, T$_{hot}$ and T$_{cold}$, in order to generate a stationary thermal conduction regime.  The heat flux $\mathbf{J_i}$($\mathbf{J_r}$) injected into (removed from) the sample is calculated as the numerical time derivative of the energy injected into (removed from) the sample per unit area\cite{melis2015barbarino}. Heat transport is here one dimensional and takes place along the $\textit{z}$ axis. From now on the heat flux will therefore be casted in scalar form. When steady state is reached, $|\mathbf{J_i}|=|\mathbf{J_r}|=|J|$, $J$ being the steady state heat flux. 
The steady state temperature spatial profile $T(z)$ is calculated starting from the atomic kinetic energy via the equipartition theorem. The steady state temperature gradient $dT/dx$ is hence retrieved. With these two ingredients at hand the thermal conductivity, $\kappa$, is finally obtained by applying the scalar version of Fourier law
\begin{equation}
J=-\kappa\frac{dT}{dz}.
\label{fourier01}
\end{equation}
The same method can be applied to interfaced systems to obtain the TBR. When a stationary state is reached, a temperature drop $\Delta T$=$T_{ih}$-$T_{ic}$ at the interface can be observed, due to the presence of a TBR, $T_{ih}$ and $T_{ic}$ being the temperatures on the hot and cold side of the interface respectively. $R_{ph}$ is calculated according to 
\begin{equation}
R_{ph}=\frac{\Delta T}{|J|}.
\label{itr}
\end{equation}

\subsubsection{Sample preparation}
\label{Sample preparation}
Al$_2$O$_3$ can be found in different crystalline phases, among which $\alpha$-Al$_2$O$_3$ (corundum) is
the most stable one at ambient conditions. We simulated this phase with the heat flux flowing along the (0001) direction.

In many crystalline systems\cite{sellan2010,dettori2014}, the lattice thermal conductivity, $\kappa_{ph}$, depends on the sample length, $L_z$, along the heat flux direction. This is because, if $L_z$ is shorter than the maximum phonon mean free path $\lambda_{max}$, phonons with mean free path in the range $L_z<\lambda<\lambda_{max}$ will not contribute to $\kappa_{ph}$. Since the experimental setup involves bulk Al$_2$O$_3$ and thin Al films, we need to estimate the minimum Al$_2$O$_3$ thickness which is able to appropriately mimic the thermal behaviour of bulk sapphire. We have thus generated  Al$_2$O$_3$ samples of different length, $L_z$, and we have calculated their lattice thermal conductivity. It has been shown\cite{puri2011} that $R_{ph}$ is substantially unaffected by the dimensions of the samples in the directions perpendicular to the heat flux, and thus for computational convenience we have  chosen to fix the lateral size of our samples to $L_x$=2.47 nm and $L_y$=2.86 nm. $L_z$ was varied between 20 and 200 nm. 

In general, the overall thermal conductivity $\kappa$ is given by the sum of lattice and electronic contributions, $\kappa=\kappa_{ph}+\kappa_e$. For thermal insulators, such as Al$_2$O$_3$, the electronic contribution is however negligible, and thus $\kappa\simeq\kappa_{ph}$. From now on, we will use $\kappa$ when referring to the conductivity of Al$_2$O$_3$, while we will distinguish $\kappa_{ph}$ and $\kappa_e$ when addressing Al.

\begin{figure}[h!]
\includegraphics[width=0.5\textwidth]{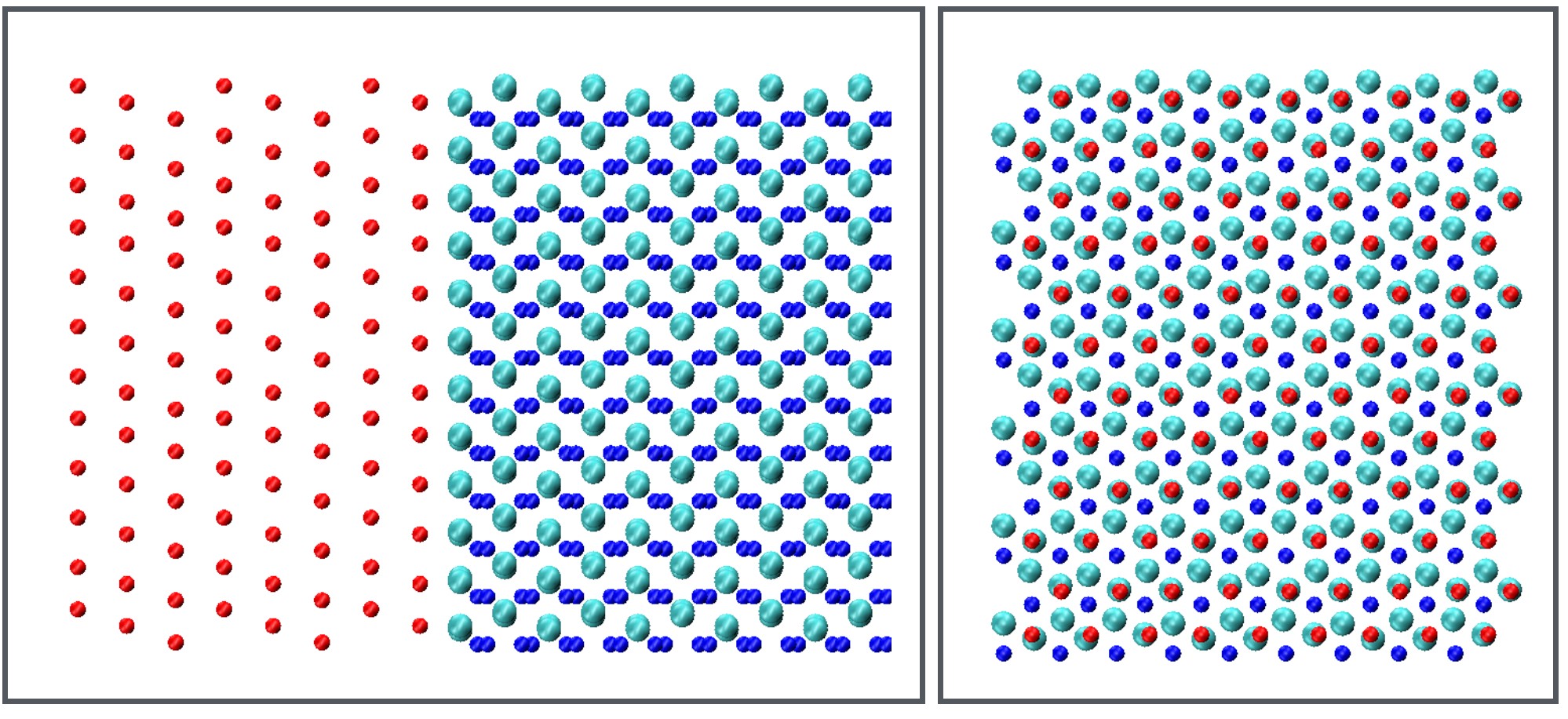}
\caption{Model for the Al/Al$_2$O$_3$ interface. Cyan represents oxygen, blue corresponds to Al atoms in sapphire and red  to Al in the metal slab. Left (right) panel shows the side (top) view.}
\label{crys}
\end{figure}

As far as concerns the Al/Al$_2$O$_3$ interface, among the possible models proposed\cite{vermeersch1990,vermeersch1995,medlin1997,siegel2002,pilania2014,mei2015} we choose the one suggested by Mei et al.\cite{mei2015} that was found to be the most stable for the Streitz-Mintmire potential. In detail, the orientation relationship between Al and sapphire is $[\bar{1}10](111)_{\mathrm{Al}}||[10\bar{1}0](0001)_{\mathrm{Al_2O_3}}$, the 
sapphire is O-terminated, and the Al atoms are located on top of the O atoms of the sapphire surface. 
Al/Al$_2$O$_3$ interfaces were constructed by coherent lattice approximation following Ref.~[\!\!\citenum{mei2015}] (see Figure\ref{crys}).

\subsection{Lattice thermal conductivity of Al$_2$O$_3$ and Al}
\label{Lattice thermal conductivity of pure Al$_2$O$_3$ and Al}
\begin{figure}[h!]
\includegraphics[width=0.5\textwidth]{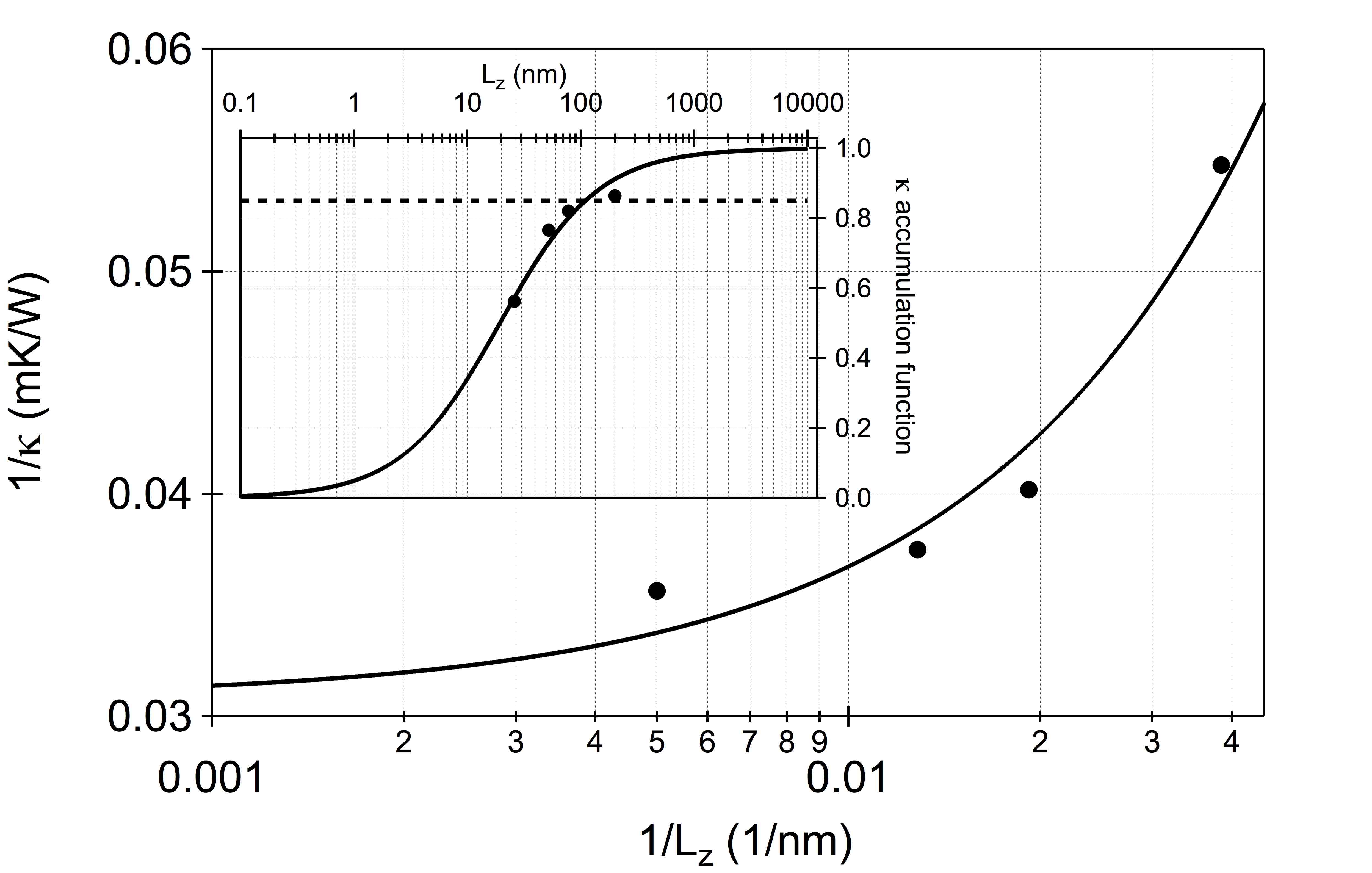}
\caption{
Black dots: reciprocal of lattice thermal conductivity of pure Al$_2$O$_3$ vs reciprocal Al$_2$O$_3$ length.
Black continuous line: fitting function $1/\kappa(L_z) = \left(1/\kappa^{\infty}\right)\left[1+\left(\lambda/L{_z}\right)\right]$ from which $\kappa^{\infty}$ is extrapolated.
Inset. Black dots: $\kappa$ accumulation function, $\kappa(L_z)/\kappa^{\infty}$, for pure Al$_2$O$_3$ vs Al$_2$O$_3$ length (top-axis). Black continuous line is the function $\kappa(L_z)/\kappa^{\infty} = \left[L_{z}/\left(\lambda + L_{z}\right)\right]$.
Grey dashed horizontal line represents 85\% of the accumulation function. 
In both graphs the abscissa are in log scale.
}
 \label{k_kaccum_sapph}
\end{figure}

The room temperature thermal conductivity of pure Al$_2$O$_3$ is calculated for 20 nm $\leq L_z\leq$ 200 nm. In order to estimate the value of $\kappa$ for $L_z=\infty$ we use the usual $1/\kappa$ vs $1/L_z$ linear extrapolation procedure\cite{sellan2010}, obtaining ${\kappa^{\infty}}$=32.5 Wm$^{-1}$K$^{-1}$, in excellent agreement with the experimental value\cite{cahill1998} of 35 Wm$^{-1}$K$^{-1}$. The results are reported in Fig.\ref{k_kaccum_sapph}. 

The inset of the same figure shows the corresponding accumulation function, defined as the ratio $\kappa(L_z)/{\kappa^{\infty}}$ between the lattice thermal conductivity calculated for a simulation cell with length
$L_z$ and the corresponding extrapolated value for $L_z$=$\infty$.
The quantity $\kappa(L_z)/{\kappa^{\infty}}$ gives the contribution to lattice thermal conductivity provided by phonons with  mean free path $\lambda$ up to $L_z$.
From the behaviour of the accumulation function it is apparent that phonons with $\lambda\leq$200 nm contribute to 85\% of the thermal conductivity (see dashed horizontal line). Furthermore, for grater sample lengths the accumulation function flattens out. A sample length of 200 nm is a good brake-even value beyond which the thermal conductivity may be thought as having reached its asymptotic value. Longer samples would lead to marginal improvements in the thermal conductivity value while requiring prohibitive calculation times. Our samples will thus consist of a $\alpha$-Al$_2$O$_3$ block of size 2.47x2.86x200 nm$^3$ in the $x$, $y$ and $z$ directions, respectively, interfaced to a fcc Al block with same $x$ and $y$ sizes and variable length along the heat flux direction $z$. From now on we will address the sapphire length $L_z$ simply as $L$, the lateral dimensions $L_x$ and $L_y$ remaining unchanged throughout the paper.

Concerning the room temperature lattice thermal conductivity of pure Al, k$_{ph}$, we find a very weak dependence on L$_{Al}$ in the range 26 to 100 nm, with a value of k$_{ph}\sim$7 W/mK. This is in agreement with recent {\it ab initio} calculations on phonon and electron transport properties of Al, showing that phonons with mean free path smaller than 20 nm contribute to 90\% of the lattice thermal conductivity\cite{mcgaughey2016}. We also notice that our NEMD value is in very good agreement with the {\it ab initio} one, the latter being $\sim$6 W/mK[\!\!\citenum{wang2016JAP}].

\subsection{Thermal Boundary Resistance}
\label{Thermal Boundary Resistance}
The first studied system corresponds to an interface between a 200 nm thick Al$_2$O$_3$ sample and a 60 nm thick Al film and contains $\sim$2$\times$10$^5$ atoms . The system has been first relaxed with a conjugated gradient minimization, followed by a low temperature (1 K) short dynamics (5 ps). 
Atomic partial charges of the whole system have been calculated during the conjugated gradient minimization run.

\begin{figure}[h!]
\includegraphics[width=0.5\textwidth]{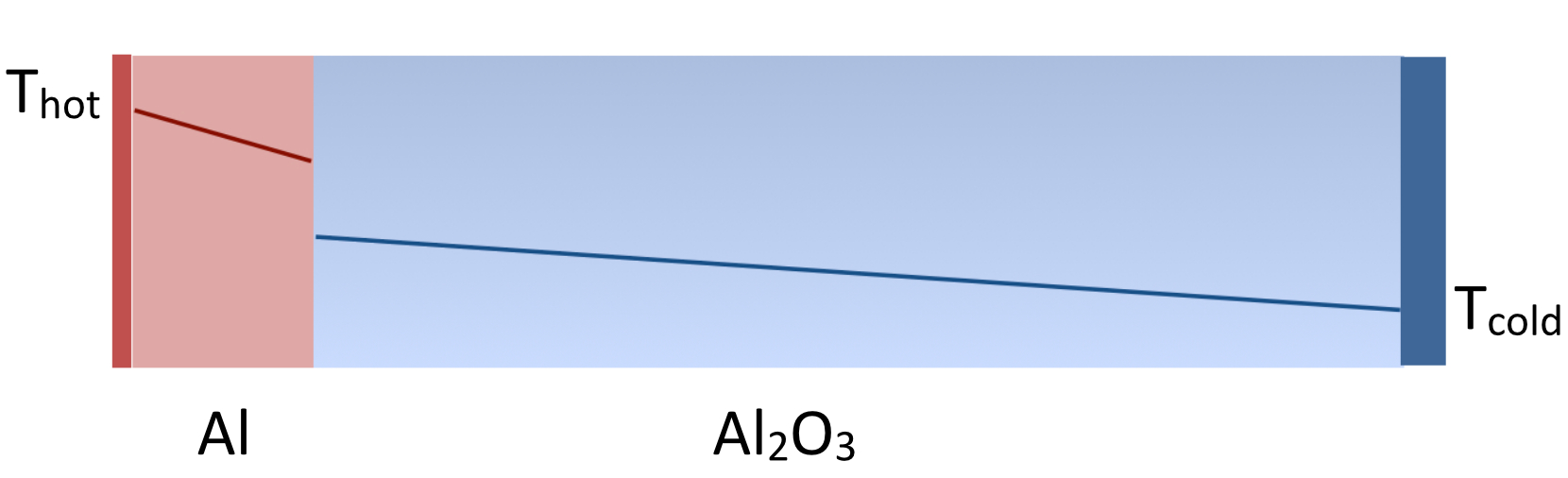}
\caption{Schematic representation of the NEMD setup (not to scale). Aluminium is heated by the thermostat set at T=T$_{hot}$, while sapphire is kept at lower temperature by the thermostat set at T=T$_{cold}$.}
 \label{schema1}
\end{figure}
After relaxation, the system has been coupled to two thermostats of size equal to 12 nm and 40 nm for the Al and Al$_2$O$_3$ sides respectively (corresponding to 20$\%$ of the metal and insulator region extents).
This situation is schematically illustrated in Fig.\ref{schema1}.  We choose T$_{hot}$=350 K and T$_{cold}$=250 K.
For this system steady state is reached after 4.5 ns and the simulation runs for further 1 ns, during which $J$ and  $\Delta T$ are calculated. The obtained value for $R_{ph}$=1.35 m$^2$K/GW.
\subsubsection{A strategy to reduce the computational workload}
\label{A strategy to reduce the computational workload}

Since the simulations described in the previous section have been in fact very computer-intensive, in view of further calculations we developed a strategy aimed at reducing the overall system size, still predicting correct TBR values. The idea is to perform calculations on a shorter model system while preserving the same $R_{ph}$.
This is achieved by considering that $R_{ph}$ only depends on $J$ and $\Delta T$ at the interface, which, in turn, depends on T$_{ih}$ and T$_{ic}$. We proceed by keeping the Al side unaltered while varying the sapphire substrate extension, $L$, and the cold thermostat temperature, $T_{cold}$, under the constraints of constant values of $T_{ic}$ and $J$.

Upon reducing the thickness of the sapphire slab from $L$ to $L'$, the thermal conductivity on the sapphire side diminishes from $\kappa$ to $\kappa'$ due to the size effect described in Sec.\ref{Sample preparation}. Requiring $T_{ic}$ to remain constant, $T_{cold}$ is left as the only quantity that may be varied in order to control the heat flux addressed in Fourier law. The thermostat temperature in the reduced slab, $T'_{cold}$, is then chosen enforcing the heat flux to be independent of slabs length:
\begin{equation}
-\kappa'\frac{\left(T_{ic}-T'_{cold}\right)}{L'}=-\kappa\frac{\left(T_{ic}-T_{cold}\right)}{L}
\label{eqred}
\end{equation}
yelding
\begin{equation}
T'_{cold}=T_{ic}+\frac{\kappa}{\kappa'}\frac{L'}{L}(T_{cold}-T_{ic})
\label{eqred2}
\end{equation}
$\kappa'$, $L'$, $T_{ic}$ and $T_{cold}$ being fixed.

\begin{figure}[h!]
\includegraphics[width=0.5\textwidth]{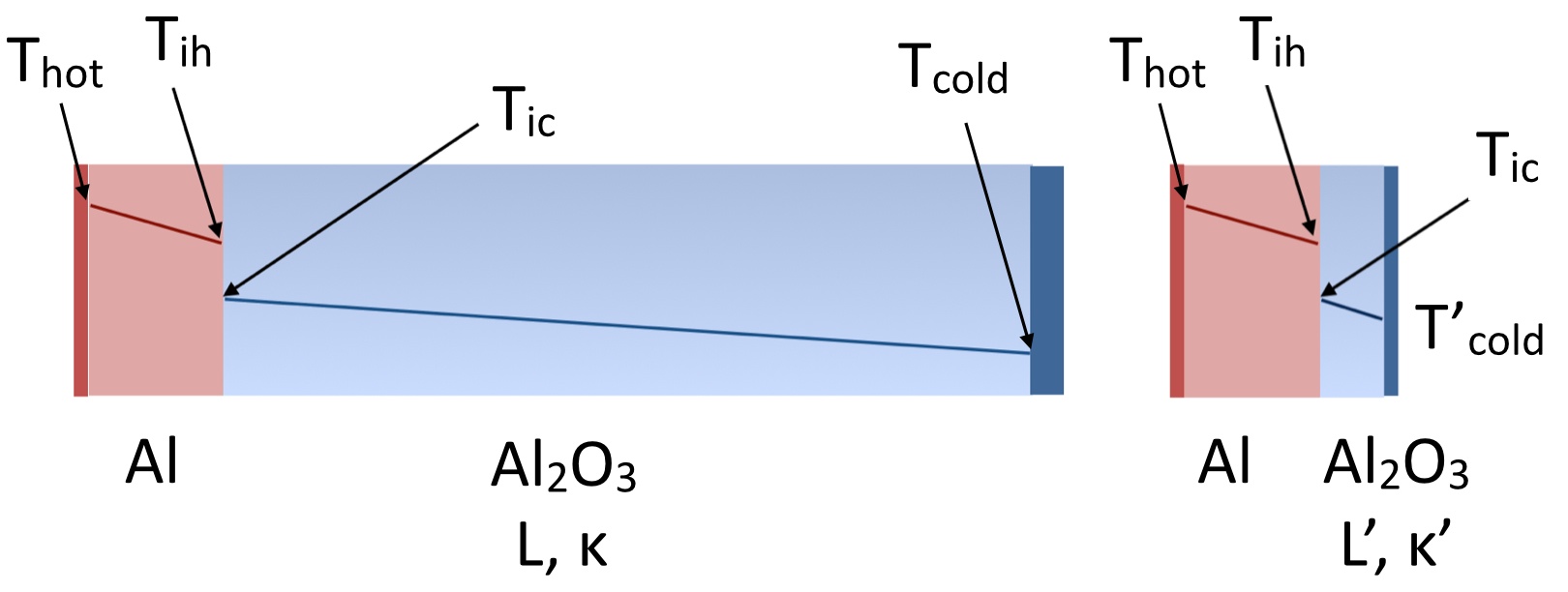}
\caption{Illustration of the sapphire thickness reduction strategy.}
 \label{riduzionefig}
\end{figure}

The procedure is illustrated in Fig.\ref{riduzionefig}, where the two equivalent systems are represented.

We have thus reduced the sapphire thickness of our sample from $L=$200 nm to only $L'=$26 nm. For the former case calculations yield $\kappa$=28.05 Wm$^{-1}$K$^{-1}$, whereas for the latter the value $\kappa'$=18.25 Wm$^{-1}$K$^{-1}$ is found. The value $T_{ic}$ is fixed at 284 K, as obtained for the sample caharacterized by $L$, $\kappa$ and $T_{cold}$. These values, inserted in Eq. \ref{eqred2},  yield $T'_{cold}$=276.2 K. 
The calculated value for $R_{ph}$ with these new parameters is 1.33 m$^2$K/GW, that is, within the numerical error, equal to the value calculated adopting the thicker substrate. This strategy allowed us to reduce the computational time to one fifth with respect to the previous case. All the following calculations are performed with the equivalent sapphire length $L'=$26 nm.

The present length-reduction strategy may be adopted on any interface system providing a substantial gain in computational time.
\subsubsection{Dependence on film thickness and interface temperature}
\label{Dependence on film thickness and interface temperature}
\begin{figure}[h!]
\includegraphics[width=0.5\textwidth]{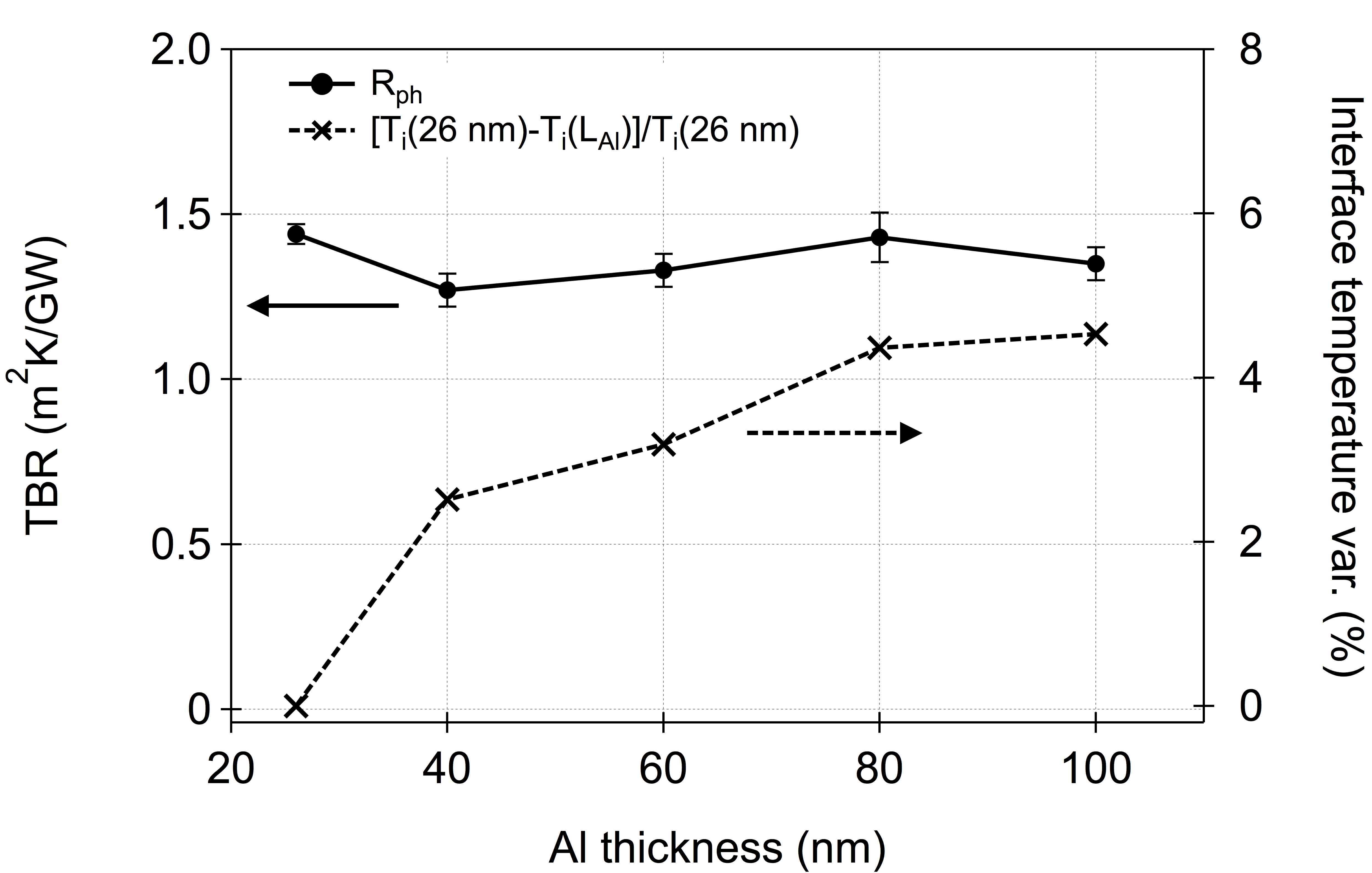}
\caption{Thermal boundary resistance (left axis) and relative interface temperature variation (right axis) vs Al thickness.}
 \label{rvsl}
\end{figure}

In order to evaluate the dependence of the TBR on metallic films thicknesses commonly encountered in TR-TR experiments, $R_{ph}$ is here calculated for an Al thicknesses $L_{Al}$ in the range 20-100 nm, keeping the hot and cold thermostats at $T_{hot}$=350 K and  $T'_{cold}$=276.2 K respectively and adopting a 26 nm thick sapphire slab ($L'=$26 nm). The results are reported in Fig.\ref{rvsl}. The values of $R_{ph}$ are constant irrespective of the Al thickness, see left axis in Fig.\ref{rvsl}. By augmenting $L_{Al}$ one potentially reduces the interface temperature $T_{i}$, where $T_{i}$=$(T_{ih}+T_{ic})/2$. The question then arises as to wether $R_{ph}$ remains constant because of a combined effect of increasing $L_{Al}$ and decreasing $T_i$. The percentage variation of $T_{i}$ with respect to the value calculated for a 26 nm thick Al film, [$T_{i}$(26 nm)-$T_{i}$($L_{Al}$)]/$T_{i}$(26 nm), is reported against $L_{Al}$ on the right axis of Fig.\ref{rvsl}, the maximum interface temperature variation amounting to 4$\%$ only. The interface temperature remains basically constant while the film thickness is varied by a factor of four, thus ruling out the scenario brought about by the aforementioned question. This proves that $R_{ph}$ is a genuine interface property not influenced by the sample length.  
Similar results have been found for other systems, e.g. for strongly coupled interfaces between crystalline silicon and amorphous silica\cite{li2012}.

\begin{figure}[h!]
\includegraphics[width=0.5\textwidth]{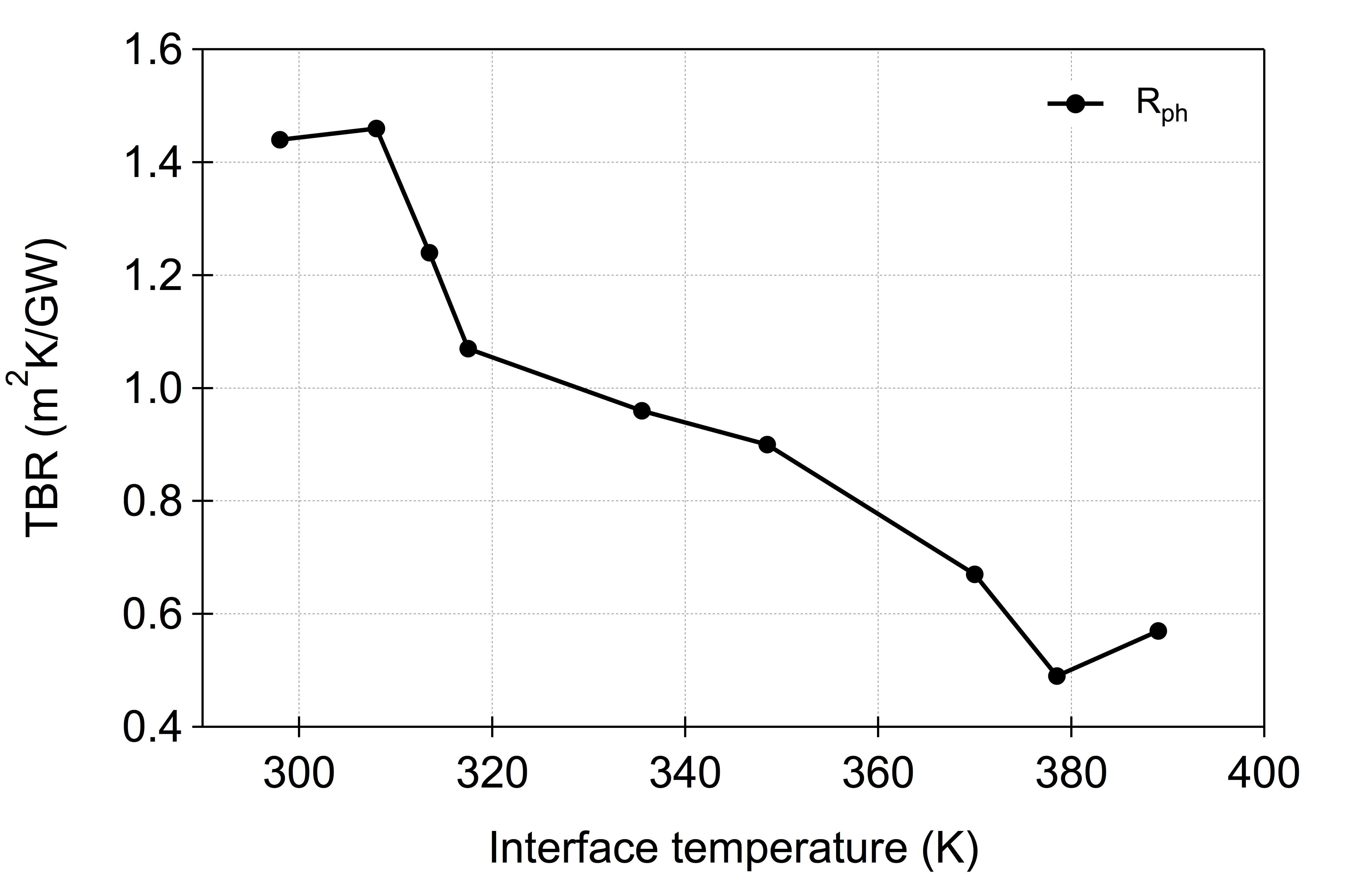}
\caption{TBR vs interface temperature.}
 \label{rvst}
\end{figure}

We now inspect the temperature dependence of $R_{ph}$ in the range 300 K $< T_{i} <$ 400 K, the latter being the interface temperature range reached with typical lasers parameters used in TR nanocalorimetric experiments performed at room temperature \cite{Nardi_NL2011, Banfi_APL2012, Banfi_PRB2010, Giannetti_PJIEEE2009}. The interface temperature $T_{i}$ is varied by changing the thermal bath temperature $T_{hot}$ on the film side while keeping $T'_{cold}$=276.2 K. For sake of computational convenience, we investigate the thinner film sample, $L_{Al}$=26 nm. Figure \ref{rvst} reports the values of $R_{ph}$ versus $T_{i}$. $R_{ph}$ decreases with increasing $T_{i}$. A similar trend has been observed experimentally for the Al/Al$_2$O$_3$ interface\cite{norris2007} as well as for other kinds of interfaces\cite{Banfi_APL2012,rurali2016}, both theoretically\cite{li2012,sellan2010} and experimentally\cite{costescu2003,lyeo2006,hopkins2009}. Such a trend has been attributed to  the increase of the anharmonicity of the atomic interactions with temperature\cite{li2012}. 
The correct scaling of TBR with temperature is a strong consistency check further confirming the validity of the adopted computational approach. 

A comment is here due regarding the eventual role on the TBR of direct coupling of metal electrons and insulator phonons at the interface. Our model solely accounts for the contribution of lattice vibrations to the thermal transfer across the interface. The underlying idea is that electrons cannot flow into the insulator side and therefore should not appreciably contribute. The question then arises as whether, in the absence of electrons flow into the substrate, metal electrons may efficiently couple to the substrate lattice vibrations directly at the interface. This point and its implications on the TBR are a debated issue\cite{Lombard2014,sadasivam2015}.  Let assume the latter scenario. The metal electrons \textit{directly} couple to nearby phonons at the insulator side, contributing a heat flux $R_{e}^{-1}$($T_{ih,e}-T_{ic}$), where $T_{ih,e}$ is the electrons temperature at the metal side of the interface and $R_{e}$ is the electronic thermal boundary resistance.
This energy flux is in \textit{parallel} with that one carried by phonons, which is addressed in NEMD simulations. The electronic thermal boundary conductance, $R_{e}^{-1}$, is to be added to the lattice one, $R_{ph}^{-1}$. In the case of the Al/Al$_{2}$O$_{3}$ interface, and within the frame of Sergeev model\cite{sergeev1998}, one obtains $R_{e}^{-1}$=0.64 GWm$^{-2}$K$^{-1}$ yielding a value of TBR=$R_{ph}R_{e}/(R_{ph}+R_{e})$ $\sim$ 0.75$\times R_{ph}$. Sergeev model tends to overestimate the electron thermal conductance \cite{Lombard2014}, for this reason an even smaller correction to the TBR is expected. Experimental evidences, obtained engineering the interface quality, also suggest the phonon channel as the dominant heat transfer mechanism across the interface \cite{stoner1993}. For these reasons in the following we will stick to the original ansatz, forgoing the role of electron-phonon coupling across the interface and identifying the TBR with $R_{ph}$.
\section{Transient nanocalorimetry}
\label{Transient nanocalorimetry}
Knowledge of the material thermal parameters, as obtained from atomistic simulations, now allows to theoretically address the impulsive thermal dynamics occurring in time-resolved all-optical calorimetry and the associated TBR retrieval process from transient thermo-reflectivity traces.

In a typical TR-TR experiment a thin metallic film is placed in thermal contact with a substrate serving as a thermal bath. An ultrafast laser pump beam delivers an energy density $\Delta$$U_{V}$ to the film triggering an impulsive thermal dynamics (\textit{excitation}). The carriers thermal dynamics affects the temperature dependent dielectric constant, ultimately resulting in a transient sample reflectivity $\mathfrak R(t)$. The sample' s temperature relaxation to the substrate is accessed measuring the transient reflectivity variation, $\Delta \mathfrak R(t)$, via a time-delayed laser probe beam (\textit{detection}). The time-delay, $t$, is taken with respect to the instant of the pump beam arrival and $\Delta \mathfrak R(t)$ = $\mathfrak R(t)$-$\mathfrak R_{0}$, where $\mathfrak R_{0}$ is the sample reflectivity in the absence of the excitation process (or equivalently $\mathfrak R_{0}$=$\mathfrak R({\infty})$ the reflectivity of the sample once relaxed back to equilibrium). 
Fitting $\Delta \mathfrak R(t)$, with the TBR as the fitting parameter, allows retrieving the TBR itself. 
It is therefore essential to link (a) $R_{ph}$ to the carriers thermal relaxation triggered in the \textit{excitation} process (b) the carriers thermal relaxation to the optical reflectivity variation $\Delta \mathfrak R(t)$, as acquired in the \textit{detection} process.  

\subsection{Excitation process: impulsive thermodynamics}
\label{Excitation process: impulsive thermodynamics}
\subsubsection{Problem formulation}
\label{Problem formulation}
The physics is addressed considering an Al thin film, 26 nm thick, deposited on a sapphire substrate, 100 $\mu$m thick. The overall sample is assumed of cylindrical shape, the diameter being in the mm range. The cylindrical sample cross-section is schematically reported in inset of Fig.\ref{TvsDelay}.  Insulating boundary conditions are applied on the top and lateral boundaries whereas the sapphire bottom boundary (z= -100$\mu$m) is kept at a constant temperature (as would be the case for a substrate adhering to a temperature controller).\\
Let us consider a scenario where the sample is excited by a single laser pulse - 5 nJ per pulse, 120 fs time duration at Full Width Half Maximum (FWHM), 780 nm central wavelength, 55 $\mu$m spatial extension at FWHM. The laser probe beam diameter in real experiments is much smaller than the pump one so as to ensure investigation of an homogeneously excited area. The thermal dynamics will be here discussed in the probed area. For this reason, and for sake of simplicity, in the present discussion the equations will be casted as if the problem was one-dimensional (1D). In the actual numerics the equations were solved in vectorial form to properly account for boundary conditions on the lateral sample frontiers and the laser spatial gaussian beam profile. It was then checked that, within the probed area, the problem solutions were in fact 1D. Within the frame of the two temperature model (TTM) \cite{Kaganov1957} and Fourier law, energy balance within the volume of the metal film reads:
\begin{eqnarray}
&&C_{el}(T_{el})\frac{\partial T_{el}}{\partial t}=P_{p}(z,t)-G(T_{el}-T_{ph})+\kappa_{e}\frac{\partial^{2} T_{el}}{\partial z^{2}}\label{TTM_el}\\
&&C_{ph}\frac{\partial T_{ph}}{\partial t}=G(T_{el}-T_{ph})+\kappa_{ph}\frac{\partial^{2} T_{ph}}{\partial z^{2}}
\label{TTM_ph} 
\end{eqnarray} 
 where $C_{ph}$=2.48$\times$10$^6$ Joule/m$^{3}$K [\!\!\citenum{Cverna2002}], and $C_{el}$=$\gamma_{e}T_{e}$ are the film phononic and electronic specific heat per unit volume respectively, $\gamma_{e}$ = 95 Jm$^{-3}$K$^{-2}$ is the Sommerfeld parameter for Al calculated starting from the data provided in Ref.~[\!\!\citenum{Allen1987PRB}] and in good agreement with the value reported in Ref.~[\!\!\citenum{lin2008}], $G$=2.45$\times$10$^{17}$ W/m$^{3}$K is the electron-phonon coupling constant\cite{hostetler99}, $\kappa_e$=230 W/mK is the Al electronic thermal conductivity \footnote{The value $\kappa_{Al}$=$\kappa_{e}$+$\kappa_{ph}$ with $\kappa_{Al}$=237 W/mK from Ref.~[\!\!\citenum{Ashcroft}], whereas $\kappa_{ph}$=7 W/mK is obtained from NEMD calculations in Section \ref{MD}, hence $\kappa_{e}$=230 W/mK}, $\kappa_{ph}$= 7.12 W/mK as calculated in Section \ref{MD} and $P_{p}$ is the profile of the pulsed power per unit volume absorbed by the film electrons. \footnote{The actual numerics has been implemented assigning the proper symmetry to $P_{p}$, $P_{p}$=$P_{p}$(r,z,t), as dictated by the pump laser gaussian profile, nevertheless, within the probed area, the problem is substantially 1D. Further details may be found in Ref.~[\!\!\citenum{Giannetti_PRB2007}]}
  $P_{p}$ has been calculated accounting for thin film effects via Fresnel relations for a multilayer system assuming $\tilde{n}_{Al}$=2.58+8.4$i$ and $\tilde{n}_{Al_{2}O_{3}}$=1.76 for the Al and Al$_{2}$O$_{3}$ indices of refraction at a wavelength of 780 nm, see Refs.~[\!\!\citenum{Rakic1998,Malitson1972}]. The film is not energetically isolated, phonons can in fact transfer energy across the interface to the sapphire substrate whereas no electrons can flow into the insulator. The heat flux at the Al-Al$_{2}$O$_{3}$ interface is ruled by $R_{ph}$, as obtained in Section \ref{MD}, through the following boundary condition:
\begin{equation}
\begin{aligned}\label{boundary_resistivity}
  -\kappa_{ph}\frac{\partial T_{ph}}{\partial z}\bigg|_{z=0+}+[T_{ph}(z=0+)-T(z=0-)]/R_{ph} &= 0\\
  -\kappa\frac{\partial T}{\partial z}\bigg|_{z=0-}+[T_{ph}(z=0+)-T(z=0-)]/R_{ph} &= 0
\end{aligned}
\end{equation}
with $T$ indicating the substrate temperature, whereas a zero flux boundary condition applies for electrons:
\begin{equation} 
-\kappa_{e}\frac{\partial T_{e}}{\partial z}\bigg|_{z=0+}=0 .
\label{electron_boundary}
\end{equation}
Given the coordinate system specified in inset of Fig.~\ref{TvsDelay}, $z=0+$ and $z=0-$ indicate the metal and substrate side of the interface respectively.
We set $R_{ph}$=1.44 m$^{2}K/GW$ (interesting enough any other value within the fluctuation range reported in Fig.\ref{rvsl} yields the same results \footnote{As discussed in subsection \ref{Thermal Boundary Resistance}, the value $R_{ph}$ does not depend on the Al thickness, the small fluctuations reported in Fig.\ref{rvsl} being due to numerical instabilities.}), and $\kappa$=${\kappa^{\infty}}$=32.5 Wm$^{-1}$K$^{-1}$ as calculated in Section \ref{MD}. We remark that $T_{ph}(z=0+)$ and $T(z=0-)$ coincide with the previously defined $T_{ih}$ and $T_{ic}$ respectively. The thermal dynamics in the sapphire volume is described by the heat equation:
\begin{equation} 
C\frac{\partial T}{\partial t}-\kappa\frac{\partial^{2}T}{\partial z^{2}}=0 
\label{sapphire_Fourier}
\end{equation}
where $C$= 3.09$\times$10$^6$ Joule/m$^{3}$K is the sapphire specific heat per unit volume\cite{weber2002handbook}.\\
As for the initial conditions we have assumed a spatially constant temperature $T_{0}$=298 K throughout the sample \footnote{As for the initial conditions we have checked the effect of the cumulative effect of the train of pulses from a cavity-dumped laser apparatus with repetition rate of 1 MHz (a cavity damped system has been assumed in order to avoid excessive cumulated heating) following Ref.~[\!\!\citenum{Banfi_PRB2010}]. The outcome is that, given the present laser parameters and materials combination, assuming a spatially constant temperature $T_{0}$=298 K is a reasonable approximation.}. 
The non-linear system of equations has been solved via the Finite Element Method covering a time-scale spanning five order of magnitudes. For ease of consultation, the materials parameters employed in the FEM simulation are summarised in Table~\ref{parameters}.
\begin{table}[h]
	\caption{Summary of materials parameters values adopted for FEM simulations. Legend:  [*] values obtained in the present work (Section \ref{MD}); [a] $\kappa_{e}$=$\kappa_{Al}$-$\kappa_{ph}$ with $\kappa_{Al}$=237 W/mK from Ref.~[\!\!\citenum{Ashcroft}] and $\kappa_{ph}$ obtained from NEMD calculations (Section \ref{MD}).}	\label{parameters}
	\begin{center}\tabcolsep=2mm	
\begin{tabular}{lccc} 
\Xhline{2.5\arrayrulewidth} 
\\ [-2.0ex]
\Xhline{2.5\arrayrulewidth} 
\\ [-1.5ex]
$\gamma_{e}$ & 95 & J m$^{-3}$K$^{-2}$ & [\!\!\citenum{Allen1987PRB}]\\ [0.5ex]
$C_{el}$ & $\gamma_{e}T_{e}$ & J m$^{-3}$K$^{-1}$ & [\!\!\citenum{Allen1987PRB}]\\ [0.5ex]
$C_{ph}$ & 2.48$\times$10$^6$ & J m$^{-3}$K$^{-1}$ & [\!\!\citenum{Cverna2002},\!\!\citenum{lin2008}]\\ [0.5ex]
$C$ & 3.09$\times$10$^6$ & J m$^{-3}$K$^{-1}$ & [\!\!\citenum{weber2002handbook}]\\ [0.5ex]
$k_{e}$ & 230 & W m$^{-1}$K$^{-1}$ & [a]\\ [0.5ex]
$k_{ph}$ & 7.12 & W m$^{-1}$K$^{-1}$ & [*]\\ [0.5ex]
$k$ & 32.5 & W m$^{-1}$K$^{-1}$ & [*]\\ [0.5ex]
$G$ & 2.45$\times$10$^{17}$ & W m$^{-3}$K$^{-1}$ & [\!\!\citenum{hostetler99}]\\ [0.5ex]
$R_{ph}$ & 1.44$\times$10$^{-9}$ & W$^{-1}$m$^{2}$K & [*]\\ [0.5ex]
$\operatorname{Re}(\tilde{n}_{Al})$ & 2.58 &  & [\!\!\citenum{Rakic1998}]\\ [0.5ex]
$\operatorname{Im}(\tilde{n}_{Al})$ & 8.4 &  & [\!\!\citenum{Rakic1998}]\\ [0.5ex]
$\operatorname{Re}(\tilde{n}_{Al_{2}O_{3}})$ & 1.76 & & [\!\!\citenum{Malitson1972}]\\ [1ex]
\Xhline{2.5\arrayrulewidth} 
\\ [-2.0ex]
\Xhline{2.5\arrayrulewidth} 
\end{tabular}
\end{center}
\end{table}
\subsubsection{Results and discussion}
\label{Results and discussion}
\begin{figure}[h!]
\includegraphics[width=0.5\textwidth]{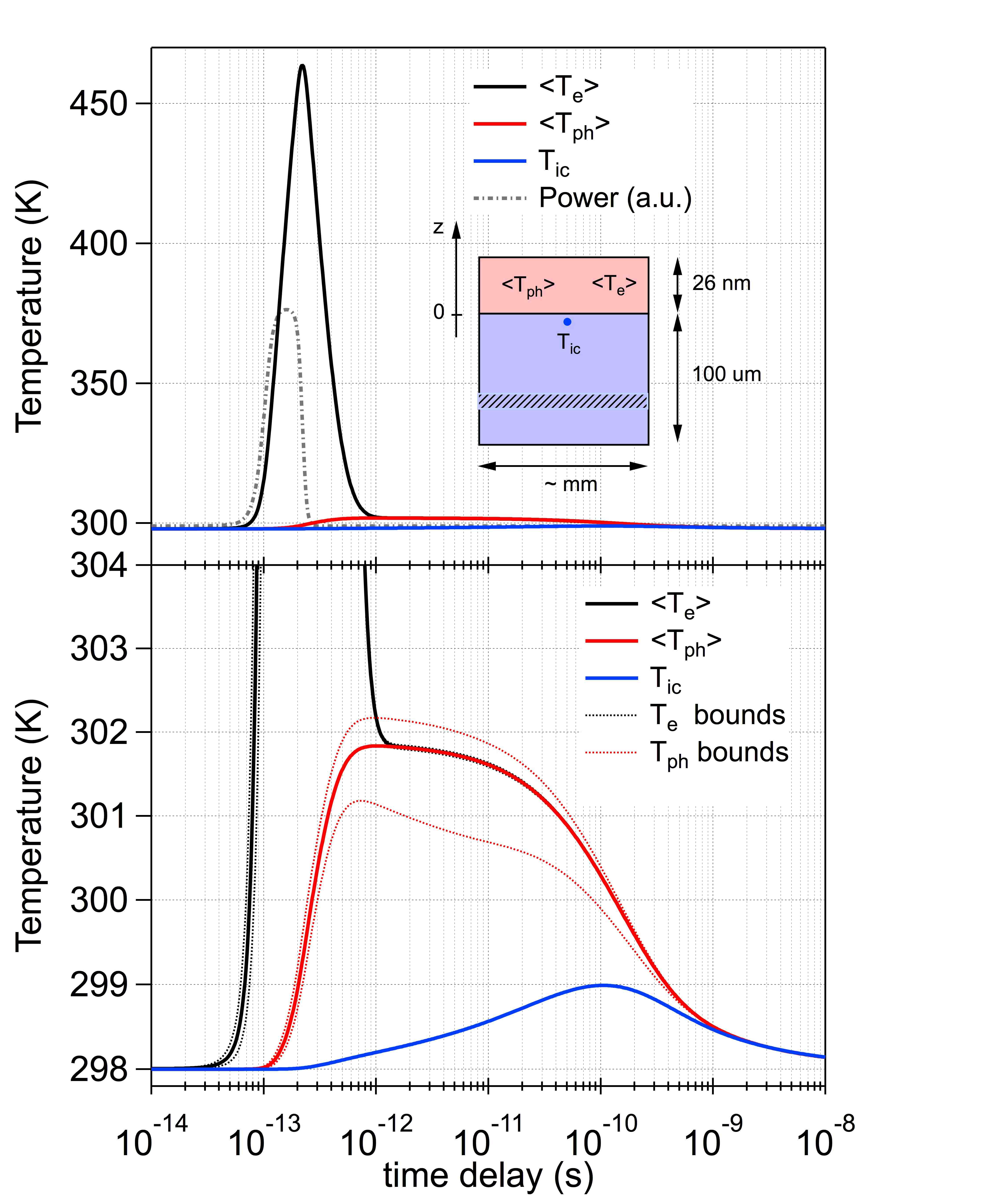}
\caption{Top panel. Average electron (black) and phonon (red) temperatures in the Al film and substrate interface temperature (blu) as a function of the time delay from the leading edge of the laser pulse (grey dash-dotted line) triggering the thermal dynamics. The averages are calculated across the film thickness. Laser pulse power time profile is reported in arbitrary units. Inset. Schematic of the simulation cell. Bottom panel. Zoom of the thermal dynamics reported in the top panel up to a temperature range of 304 K. The phonon and electron temperatures across the film thickness are bounded by the dotted red and dotted black curves respectively. The upper dotted curves are calculated at the Al film free standing surface, the lower dotted curves at the Al interface with Al$_{2}$O$_{3}$. The dotted black curves coincide with the black curve for delay longer than 100 femtoseconds, the electron temperatures being spatially constant for longer delay times. The delay time is reported in log-scale.
}
 \label{TvsDelay}
\end{figure}

A general overview of the relevant thermal dynamics, triggered by the pump laser pulse (black dash-dotted line), may be appreciated inspecting the top panel of Figure \ref{TvsDelay}. The average electrons, $\langle T_{e}\rangle$ (black line), and phonons, $\langle T_{ph}\rangle$ (red line), temperatures as calculated across the Al film thickness are therein reported together with the temperature $T_{ic}$ at the Al$_{2}$O$_{3}$ side of the interface (blu line). The evolution naturally breaks down in time steps each ruled by a specific hierarchy of processes.\\
\indent{\textit{Electron excitation step}}. The electron gas is heated up by the pump pulse on a time scale of few hundreds of femtoseconds, the maximum average electrons temperature $max\{\langle T_{e}\rangle\}$ being attained within the 120 fs pulse duration. The value $max\{\langle T_{e}\rangle\}$ is hampered by the temperature-dependent electronic specific heat per unit volume and, to a lesser extent, by the electron-phonon interaction.\\
\indent\textit{Electron-phonon thermalisation step}. The electron gas then cools raising $\langle T_{ph}\rangle$. This process continues up to a time-delay of few ps when $\langle T_{e}\rangle$ and $\langle T_{ph}\rangle$ reach the common value of 301.8 K, the dynamic being better appreciated zooming the temperature scale, as reported in bottom panel of Figure \ref{TvsDelay}. Starting from this instant one has $\langle T_{ph}\rangle$=$\langle T_{e}\rangle$ and most of the Al film energy content is stored in the phonon gas, being $C_{ph}/C_{el}\sim$10$^{2}$. The difference between the phonon and electron specific heats also accounts for the fact that $max\{\langle T_{e}\rangle\}$$\gg$$max\{\langle T_{ph}\rangle\}$. At this stage the temperature increase at the substrate interface is always rather small, amounting to $\sim$20$\%$ of its maximum excursion of 1 K, implying that, on these time scales, the Al film remains substantially isolated.\\
\indent$\textit{Film-substrate thermalisation step}$. On longer time scales the Al film thermalises with the substrate: $\langle T_{ph}\rangle$ and $\langle T_{e}\rangle$ decrease together while $T_{ic}$ attains its maximum value $\sim$100 ps after laser excitation and then decreases monotonously. $T_{ic}$ now decreases because the energy flux from the metal overlay to the substrate interface is less than the energy flux from the latter to the Al$_{2}$O$_{3}$ bulk. This process ends after 1 ns when the Al film and substrate temperatures converge to the common value of 298.5 K.\\
\indent$\textit{Proximal-bulk substrate thermalisation step}$. Starting from this instant the temperature drop across the interface is negligible as compared to the previous steps, $\Delta T$$\sim$0. The metal overlay and the substrate portion proximal to the interface thermalise together with the Al$_{2}$O$_{3}$ bulk, the thermal dynamics being ruled solely by the sapphire thermal parameters through Eq.\ref{sapphire_Fourier}.

We now focus on the $\textit{film-substrate}$ $\textit{thermalisation step}$. It is in this step that a finite value of $R_{ph}$ appreciably affects the film thermal dynamics as compared to the case of an isolated thin film where $R_{ph}$$\to$$\infty$. For this same reason the TBR is retrieved from TR-TR experiments by detecting the thermally-driven transient optical changes taking place during this step. The bottom panel of Figure \ref{TvsDelay} shows that $T_{ph}$ varies across the film thickness, its values being bound by the red dotted curves. The low temperature bound coincides with $T_{ih}$ and occurs at $z=0+$, i.e. at the metal side of the interface, whereas the high temperature bound is found at the Al film surface, z = 26 nm. On the contrary $T_{e}$ is spatially homogeneous across the film thickness. The bottom panel of Figure \ref{TvsDelay} shows that the range of $T_{e}$ values, bound by the black dotted lines, has negligible spread and $T_{e}$=$\langle T_{e}\rangle$ across the entire film thickness. These evidences also show that electrons and phonons thermalisation is far for being complete and continues throughout the entire step. The thermalisation therein addressed refers to the fact that the electrons and phonons temperatures \textit{spatial averages} have the same value.

\begin{figure}
\includegraphics[width=0.5\textwidth]{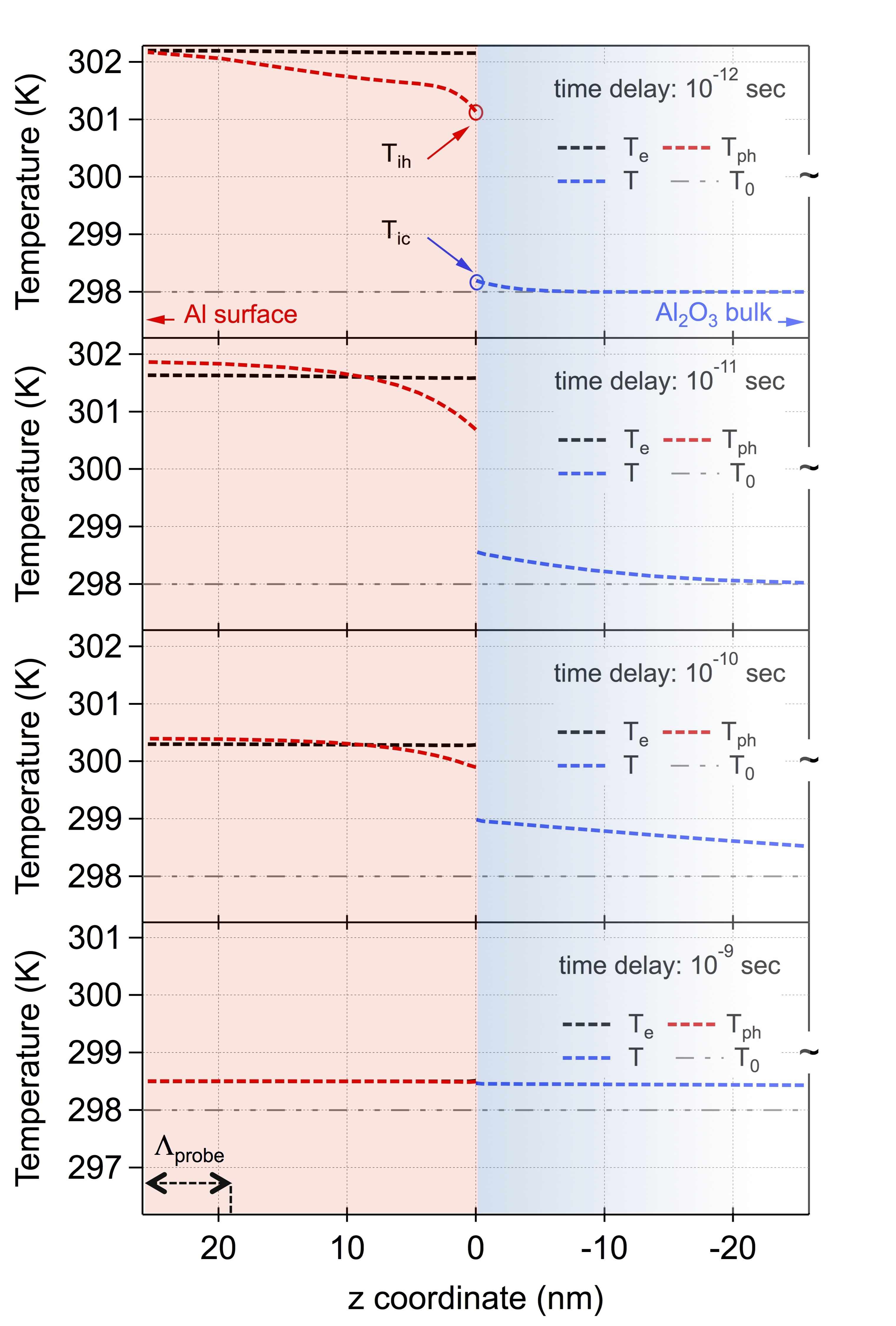}
\caption{Carriers temperatures across a portion of the sample thickness (z$\in$[-26 nm, 26 nm]) at given time delays (in ascending order from 1 ps -top panel-  to 1 ns -bottom panel) falling within the $\textit{film-substrate thermalisation step}$. The coordinate system and the colours assigned to the two materials follow the criteria adopted in the inset of top panel of Figure \ref{TvsDelay}. Temperatures profiles: metal electrons (black dashed line), metal phonons (red dashed line), substrate lattice (blu dahsed line), initial temperature (gray dash-dotted line). At greater depths, not reported in the present figure, the substrate temperature attains $T_{0}$=298 K. The laser probe penetration depth $\Lambda_{probe}$ is schematically reported in the bottom panel.}
 \label{fix_time_var_z}
\end{figure}

Further physical insight is provided by inspection of the spatial temperatures distribution across the sample depth calculated at specific delay times limiting or falling within the $\textit{film-substrate}$ $\textit{thermalisation step}$, see Figure \ref{fix_time_var_z}. One ps after the pump pulse arrival $T_{e}$$\sim$$T_{ph}$ within the first 6 nm from the film surface whereas deeper into the Al film $T_{e}$$\ge$$T_{ph}$. The substrate temperature $T$(z) remains substantially unaltered. The phonon temperature drop across the interface is $\Delta T$=2.94 K. For a time delay of 10 ps one finds $T_{ph}$$\gtrsim$$T_{e}$ within the first 16 nm from the Al surface. This small temperature overshoot is due to the fact that electrons are loosing energy not only to the phonon gas but also via diffusion to electrons located deeper into the film. Furthermore, the minute value of $\kappa_{ph}$, as compared to $\kappa_{e}$, does not allow phonons in proximity of the surface to loose energy efficiently via diffusion to other film phonons. The interplay of these two effects leads to the above mentioned overshoot. Deeper into the substrate $T_{ph}$ decreases rapidly and falls below $T_{e}$ by as much as $\sim$ 0.9 K at the interface. Phonons in proximity of the interface, as opposed to phonons residing close to the surface, have an efficient energy loss pathway towards the phonons of the substrate thus preventing the phonon-electron temperature overshoot. The energy flux across the interface $J$(t,z=0), from now on addressed as $J_{bd}$(t), effectively raises $T_{ic}$ above the $T_{0}$ baseline whereas the substrate bulk is not yet affected being $T$(z$\le$-20nm)= $T_{0}$. Across the interface one finds $\Delta T$=2.12 K. After 100 ps the electrons and phonons temperatures spread across the Al depth diminishes. $J_{bd}(t)$ continued transferring energy to the substrate, resulting in a $\Delta T$=0.91 K. Diffusion now plays a considerable role also in the substrate as may be appreciated inspecting $T$(z) up to a depth of 26 nm from the interface. Furthermore, at this instant $T_{ic}$ attains its maximum value and for longer time delays it will decrease, see blue curve in bottom panel of Figure \ref{TvsDelay}. This is due to heat diffusion becoming more efficient in dissipating energy from the cold side of the interface as compared to $J_{bd}$(t) pumping energy into it. The snapshot taken 1 ns after laser excitation shows the film and the proximal portion of the substrate at roughly the same temperature and jointly thermalizing with the substrate bulk. The thermal dynamics for longer delay times is ruled by the Al$_{2}$O$_{3}$ lattice thermal conductivity $\kappa$ via Equation \ref{sapphire_Fourier}. One may thus expect a power-law scaling for the temperature profiles for long time delays.

The time-dependent spatial detachment between $T_{e}$ and $T_{ph}$ is ruled by the interplay of thermal parameters and film thickness. The temperature detachment diminishes upon increasing $G$ and/or $R_{ph}$, the latter consideration being of the utmost importance in view of extracting $R_{ph}$ from experimental time-resolved traces.

\subsection{Detection process: TBR from Time Resolved Thermo-Reflectance.}
\label{Detection process: TBR from Time Resolved Thermo-Reflectance}
Knowledge of the carriers temperature evolution, as acquired via FEM modelling of the \textit{excitation} process, allows to link the optical reflectivity changes $\Delta \mathfrak R(t)$, taking place in TR-TR experiment, to the carrier thermal dynamics, and, ultimately, to the TBR.

Again, the focus is on the $\textit{film-substrate}$ $\textit{thermalisation}$ $\textit{step}$ occurring on time scales exceeding few ps. The reflectivity variation reads\cite{norris2003,Giannetti2016}:
\begin{eqnarray}
  \label{thermoreflectivity}
 \begin{split}
 \Delta \mathfrak R(t)= a\Delta T_{e}\left(t,z>L_{Al}-\Lambda_{probe}\right)\\
  + b\Delta T_{ph}\left(t,z>L_{Al}-\Lambda_{probe}\right)
 \end{split}
\end{eqnarray}
where $a=\partial\mathfrak R/\partial T_{e}$, $b=\partial\mathfrak R/\partial T_{ph}$, $\Delta T_{e}(t,z)$=$T_{e}(t,z)-T_{0}$, $\Delta T_{ph}(t,z)$=$T_{ph}(t,z)-T_{0}$, $\Lambda_{probe}$ is the laser probe penetration depth and $z$$>$$L_{Al}-\Lambda_{probe}$ identifies the Al depth explored by the probe laser. Eq.~\ref{thermoreflectivity} holds since both $T_{e}(t,z)$ and $T_{ph}(t,z)$ are substantially constants within a depth $\Lambda_{probe}$$\sim$ 7 nm from the Al surface for any time delay $t$. The spatial constancy strictly holds for $T_{e}(t,z)$ whereas a minimal $z$ dependence sets in for $T_{ph}\left(t,z>L_{Al}-\Lambda_{probe}\right)$ on the sub-10 picoseconds time scale, see Fig.~\ref{fix_time_var_z}.\\
Eq.~\ref{thermoreflectivity} simplifies to $\Delta \mathfrak R(t)\propto \Delta T_{e}\left(t,z\right)=\Delta T_{e}(t)$. This is achieved upon inspection of Fig.~\ref{fix_time_var_z} showing that (a) $\Delta T_{ph}\left(t,z>L_{Al}-\Lambda_{probe}\right)\simeq\Delta T_{e}\left(t,z>L_{Al}-\Lambda_{probe}\right)$, the deviation being limited to the 10 ps time-scale (b) $\Delta T_{e}(t,z)$ is actually spatially constant throughout the entire Al thickness and may thus be substituted for $\Delta T_{e}\left(t,z>L_{Al}-\Lambda_{probe}\right)$ just writing $\Delta T_{e}(t)$.
We pinpoint that $\Delta T_{e}(t)$ is a functional of $R_{ph}$, $\Delta T_{e}(t;R_{ph})$.\\
In order to correctly retrieve $R_{ph}$ from a TR-TR experiment one should therefore fit the experimental trace $\Delta \mathfrak R(t)$ with $\Delta T_{e}(t;R_{ph})$ as obtained from FEM modelling (see section \ref{Problem formulation}), taking $R_{ph}$ as the fitting parameter. The value obtained from this fitting procedure should be taken as the benchmark against which compare theoretically calculated values. 

In the case of Al on sapphire experimental values for $R_{ph}$ vary between $\sim$3.2 m$^2$K/GW~[\!\!\citenum{hopkins2011}] and $\sim$5.2 m $^2$K/GW~[\!\!\citenum{stoner1993,norris2007}], thus exceeding the value calculated in the present work \footnote{Reference \citenum{stoner1993} reports in the main text a value of $R_{ph}$$\sim$10 m$^2$K/GW, nevertheless, upon digitazing the experimental values therein reported, we retrieve 5.5 m$^2$K/GW}. We argue that the mismatch may be due to the fitting procedure adopted\cite{stoner1993,norris2007} to extract the TBR. In the TBR retrieval process the common ansatz is that, on time-scales beyond tens of ps, the electrons and phonons within the metallic thin-film (a) attain mutual thermal equilibrium, $T_{e}(t,z)$=$T_{ph}(t,z)$ $\forall$ $z$, thus allowing to define a unique thermodynamical temperature $T_{Al}(t, z)$, and (b) the Biot number $Bi$=$\left(L_{Al}/\kappa_{Al}\right)$$R_{ph}^{-1}$$\ll$1, where $\kappa_{Al}$=$\kappa_{ph}$+$\kappa_{e}$, thus allowing the metal film to be treated as a lumped thermal capacitance at temperature $T_{Al}(t)$ (no $z$ dependence) exchanging heat with the Al$_{2}$O$_{3}$ substrate at a temperature $T(t,z)$. In the seminal work of Ref.~[\!\!\citenum{stoner1993}] the fit to the data is performed allowing for two fitting parameters, $R_{ph}$ and $\kappa$. The best fit is obtained with $R_{ph}$=5.2 m$^2$K/GW and a value of $\kappa$ overestimating by a factor of four the value reported in the literature for bulk Al$_{2}$O$_{3}$. This suggests that the ansatz behind the fitting model may not be valid, a fact that clearly emerges in our simulations where the electrons and phonons temperatures at the interface remain decoupled over several time decades and the phonon temperature is spatially inhomogeneous, see Fig.~\ref{fix_time_var_z}. In order to further substantiate this point we fitted $\Delta \mathfrak R(t)$ as retrieved from our theoretical calculations, $\Delta \mathfrak R(t)$$\propto$$\Delta T_{e}(t)$, with the lumped thermal capacitance model. We allowed $R_{ph}$ as the only fitting parameter. The fit was performed on the trace $\Delta T_{e}$, as obtained from Fig.~\ref{TvsDelay} upon subtraction of the baseline temperature of 298 K. The best fit overestimates by a factor of $\sim$ 2 the theoretical value of $R_{ph}$=1.44 m$^2$K/GW used to calculate $\Delta T_{e}$. Assuming experimental values were overestimated by the same factor leads to a convergence between experimental results and the results obtained from NEMD in subsection \ref{Thermal Boundary Resistance}. 

Let us now address the physical explanation of why a fit based on the lumped thermal capacitance model overestimates the actual TBR. The TBR is defined by means of the phonons temperatures on both sides of the interface, $R_{ph}$=$\left(T_{ih}-T_{ic}\right)/J_{bd}$, whereas the probe is sensitive to the temperature dynamics taking place within a depth $\Lambda_{probe}$ from the free Al surface, $\Delta T_{e}$=$T_{e}\left(t,z>L_{Al}-\Lambda_{probe}\right)-T_{0}$. Our calculations show that the temperature entering the definition of TBR, i.e. the phonon temperature on the metal side of the interface, deviates from the temperature of the probed region. On the other hand a lumped thermal capacitance model implies a unique temperature common to both electrons and phonons throughout the entire Al film thickness leading to an overestimation of $T_{ih}$ in the expression for $R_{ph}$, and, ultimately, of the $R_{ph}$ value itself. All in all, within the lumped thermal capacitance model, the red dashed curve for $T_{ph}$ shown in Fig.~\ref{fix_time_var_z} was made to coincide with the black dashed curve for $T_{e}$ throughout the entire Al depth. On a general basis the lumped thermal capacitance model may be retrieved from our model for $R_{ph}$$\to$$\infty$. Indeed, should an interface bear an high enough TBR, the phonon temperature would be spatially homogeneous and match the electronic temperature throughout the entire metal film depth. Nevertheless, from an experimentalist point of view, the TBR is not known a-priori, hence the full thermodynamical approach should be exploited to fit the experimental TR-TR traces in order to retrieve the TBR.

\subsection{Steady-state VS transient nanocalorimetry.}
\label{Steady-state VS transient nanocalorimetry}
The results presented in the two previous subsections apply to \textit{transient} nanocalorimetry across a metal-insulating interface. Our results account for the effects of both phonons and electrons (with the exclusion of the \textit{direct} coupling mechanism between the metal's electrons and the substrate's phonons addressed at the end of Section \ref{Thermal Boundary Resistance}) and provide a protocol to access TBR from TR-TR measurements. At the light of our findings we now review literature's results in the frame of \textit{steady-state} heat transport.

In recent years extensive effort has been devoted to account for the effects of both electrons and phonons in the frame of \textit{steady-state} heat transport across a metal-insulator heterojunction. \textit{Steady-state} heat transport may be achieved for instance by placing a hot (cold) temperature reservoir in contact with the metal (insulator) end. The question then arises as to which \textit{overall} thermal resistance $R_{tot}$ one actually accesses when performing a \textit{steady-state} measurement with the scope of retrieving the TBR, given the presence of conducting electrons exchanging energy with phonons on the metal side of the interface. $R_{tot}$ encloses the contributions from all thermal resistances comprised between the two thermal leads, that is the thermal resistances within the metal's and insulator's bulk and $R_{ph}.$ It has been argued\cite{majumdar2004,wang2012,donadio2015} that, due to electron-phonon coupling, electrons may have a strong impact on $R_{tot}$. Majumdar and Reddy\cite{majumdar2004} found the analytical expression incorporating such contribution in the case of a infinitely thick metal layer in contact with an insulating substrate, suggesting that, due to electron-phonon coupling, electrons contribute a series resistance, $R_{ep}$, to the overall $R_{tot}$. Ordonez-Miranda et al.\cite{miranda2011} improved the model including the effect of the finiteness of the metal layer, an essential step forward in view of modelling actual experiments. The model describes the electrons and phonons in the metal with a \textit{steady state} version of the TTM. The time-dependent terms on the left side of Eq.s~\ref{TTM_el} and \ref{TTM_ph} are set to zero and the impulsive heating term $P_{p}$ is missing. Continuity of the heat flux across the interface involves phonons only and is achieved via Eq.s~\ref{boundary_resistivity} through \ref{electron_boundary}. The phonon thermal dynamics on the insulator side of the interface is accounted by Laplace's equation. Dirichlet (first-type) boundary conditions are enforced on the external boundaries of the system, where the metal's electrons and phonons are kept at a temperature $T_{hot}$ and the substrate's phonons at $T_{cold}$. This is at variance with the insulating boundary conditions necessary to describe the impulsively heated metal film. The model has a straightforward analytic solution as opposed to our case. The electrons, $T_{ih,e}$, and phonons, $T_{ih}$, temperatures at the metal side of the interface are found to be different. Defining $R_{tot}$=$\left( T_{hot}-T_{cold} \right)/J$, where $J$ is the \textit{steady state} heat flux across the sample, one obtains:
\begin{equation}
R_{tot}=\frac{L_{Al}}{\kappa_{e}+\kappa_{ph}}+R_{ph}+\frac{L_{Al_{2}O_{3}}}{\kappa}+R_{ep}
\label{rtotordonez}
\end{equation}
where the thermal resistance $R_{ep}$ arises because of the presence of the electron-phonon coupling and has a simple analytical expression \footnote{The explicit expression for $R_{e}$ is the fourth term on the right side of Eq.11 in Ref.~[\!\!\citenum{miranda2011}]}. $R_{ep}$ is often addressed as the electronic contribution to the TBR. This is due to the fact that the electrons and phonons temperatures decoupling is strongest at the interface, a common feature regardless of the \textit{transient} or \textit{static} nature of the problem. For sake of clarity we stress that $R_{ep}$ has nothing to do with $R_{e}$ ruling the \textit{direct} coupling mechanism between the metal electrons and the substrate phonons addressed at the end of Section \ref{Thermal Boundary Resistance}. The TBR is thus obtained measuring $R_{tot}$ in a steady state experiments and extracting $R_{ph}$ from Eq.~\ref{rtotordonez}.

This model might correctly predict $R_{tot}$ in \textit{steady-state} experiments provided that the electrons and phonons temperatures at the hot metal end can actually be set to the same value in a true experiment. This is not obvious since, in the real case, an interface (junction) might be present between the circuit connecting lead and the Al slab. However, the model cannot be used to describe the thermal dynamics occurring in TR-TR.  As a matter of fact in time-resolved nano-calorimetry a finite energy is impulsively delivered to the metal as opposed to fixing the temperature at the metal extremity. One is thus faced with a \textit{transient} as opposed to a \textit{steady-state} thermal problem. Despite this fact the \textit{steady-state} formulation has been adopted to discuss the TBR retrieved from TR-TR experiments\cite{majumdar2004,wang2012}. This is possibly due to its elegant analytical solution, explicitating in a simple formula the relevant thermal and geometrical parameters involved.
\section{Conclusions}
In the present work the Thermal Boundary Resistance in the paradigmatic 
nanometric Al film-sapphire heterojunction as been investigated by means of Non-Equilibrium Molecular Dynamics. A strategy has been devised to considerably reduce the computational burden, allowing to simulate a system of realistic size. The approach is readily transferable to other heterojunctions. The TBR was found to be 
monotonously decreasing with interface temperature in the range 300-400 K, the room temperature value being $\sim$1.4 m$^2$K/GW.
The effect of the calculated TBR on the thermal dynamics occurring in an all-optical time-resolved nanocalorimetry experiment was theoretically addressed casting the problem in terms of continuous macrophysics equations and solving them by Finite Element Methods. The electrons and phonons within the Al films were found to remain out of mutual equilibrium up to 1 ns delay-times. Whereas the electronic temperature is spatially constant throughout the metal film depth, the same does not hold for the phonons temperature, its spatial gradient attaining the maximum value in proximity of the metal-insulator interface. Knowledge of the spatio-temporal carriers dynamics allowed to link the optical reflectivity changes to the TBR, thus providing a protocol to extract the latter from real experiments. The procedure adopted in the literature to extract TBR from experimental results was revised at the light of the present findings, improving the congruency between theoretical predictions and experimental findings. 
A comparison with models addressing TBR in the frame of \textit{steady-state} heat transfer experiment was finally presented.

The present understanding goes beyond the specific system here investigated and it may be applied to other metal-insulator heterojunctions.

\begin{acknowledgments}
C.C and F.B acknowledge financial support from the MIUR Futuro in ricerca 2013 Grant in the frame of the
ULTRANANO Project (project number: RBFR13NEA4). F.B, G.F and C.G acknowledge support from Universit\`a Cattolica del Sacro Cuore through D.2.2 and D.3.1 grants. F.B and G.F acknowledge financial support from
Fondazione E.U.L.O. R.R acknowledges financial support by the Ministerio de Econom\'ia y Competitividad (MINECO) under grant FEDER-MAT2013-40581-P and the Severo Ochoa Centres of Excellence Program under Grant SEV-2015-0496 and by the Generalitat de Catalunya under grants no. 2014 SGR 301. We thank Marco Gandolfi and Daniele Scopece for enlightening discussions.
\end{acknowledgments}


\bibliography{BibliographyFra6}

\end{document}